\documentclass[3p,10pt,preprint,twoside,fleqn,sort&compress]{elsarticle}
\makeatletter
\def\ps@pprintTitle{%
   \let\@oddhead\@empty
   \let\@evenhead\@empty
   \let\@oddfoot\@empty
   \let\@evenfoot\@oddfoot
}
\makeatother

\usepackage{graphicx}
\usepackage{dcolumn}
\usepackage{bm}
\usepackage{graphicx}
\usepackage{verbatim}
\usepackage{amsmath}
\usepackage[varg]{pxfonts}
\usepackage{eepic}
\usepackage{amssymb}
\usepackage{bbold}
\usepackage{mathtools}
\usepackage[all,cmtip]{xy}

\usepackage{tikz}
\usetikzlibrary{matrix,arrows}
\usepackage{multirow}
\usepackage{bbm}
\usepackage[english]{babel}
\usepackage{stackengine}
\usepackage{scalerel}

\newcommand{\defeq}{=}
\newtheorem{theorem}{Theorem}[section]

\newtheorem{remark}[theorem]{Remark}

\DeclareMathOperator*{\argmax}{argmax}

\newcommand{\laux}{\left(}
\newcommand{\raux}{\right)}

\newcommand{\op}{\oplus}
\newcommand{\ra}{\rightarrow}
\newcommand{\lb}{\left(}
\newcommand{\rb}{\right)}
\newcommand{\pd}{\partial}
\newcommand{\dprime}{{\prime\prime}}

\newcommand{\iden}{\mathbb{1}}
\newcommand{\h}{h}
\newcommand{\Log}{\text{Log}}
\newcommand{\Exp}{\text{Exp}}

\newcommand{\sR}{\mathbb R}
\newcommand{\sN}{\mathbb N}
\newcommand{\sRp}{\sR^+_0}

\newcommand{\mcP}{P}
\newcommand{\escp}[1]{P_{#1}^{(\alpha)}}
\newcommand{\DeltaE}{\Delta E_i}

\newcommand{\Ra}{R_\alpha}

\newcommand{\Hah}{H_{\alpha}}
\newcommand{\HahInv}{H_{\alpha^{-1}}}
\newcommand{\SMaq}{SM_{\alpha,q}}
\newcommand{\SMao}{SM_{\alpha,0}}

\newcommand{\SEeq}{\stackon[-4pt]{SE}{\vstretch{1}{\hstretch{0.4}{\widehat{\phantom{\;\;\;\;\;\;\;\;}}}}}_{\alpha,r}}
\newcommand{\SEag}{SE_{\alpha,r}}

\newcommand{\Req}{\hat R_\alpha}
\newcommand{\ReqL}{\Req^{L}}
\newcommand{\ReqE}{\Req^{E}}

\newcommand{\ReqLinv}{\hat R^{L}_{\alpha^{-1}}}

\newcommand{\Pru}{\hat P_{\alpha}}
\newcommand{\PrL}{\hat P^{L}_{\alpha}}
\newcommand{\PrE}{\hat P^{E}_{\alpha}}
\newcommand{\PrLinv}{\hat P^{L}_{\alpha^{-1}}}

\newcommand{\PrEEsc}{\laux\PrE\raux^{(\alpha)}}

\newcommand{\prL}[1]{\hat p^{L}_{#1} (\alpha)}
\newcommand{\prLInv}[1]{\hat p^{L}_{#1} (\alpha^{-1})}
\newcommand{\prE}[1]{\hat p^{E}_{#1} (\alpha)}

\newcommand{\HeqL}{\Heq^{L}}
\newcommand{\HeqE}{\Heq^{E}}

\newcommand{\HeqLinv}{{\hat H}^{L}_{\alpha^{-1}}}

\newcommand{\betaRL}{\beta^{\, L}_{\alpha}}

\newcommand{\betaRLInv}{\beta^{\, L}_{\alpha^{-1}}}

\newcommand{\ReqU}{\Req}
\newcommand{\Tru}{T_{\alpha}}
\newcommand{\Fru}{F_{\alpha}}
\newcommand{\Cru}{C_{\alpha}}

\newcommand{\betaRU}{\beta_{\alpha}}

\newcommand{\betaRE}{\beta^{\, E}_{\alpha}}

\newcommand{\Fhe}{\overline F^{E}_{\alpha}}
\newcommand{\Che}{\overline C^{E}_{\alpha}}
\newcommand{\Zhe}{\overline Z^{E}_{\alpha}}
\newcommand{\betaHE}{\overline \beta^{\, E}_{\alpha}}

\newcommand{\FhlInv}{\overline F^{L}_{\alpha^{-1}}}
\newcommand{\ChlInv}{\overline C^{L}_{\alpha^{-1}}}
\newcommand{\ZhlInv}{\overline Z^{L}_{\alpha^{-1}}}
\newcommand{\betaHL}{\overline \beta^{L}_{\alpha}}
\newcommand{\betaHLAm}{\overline \beta^{L}_{\alpha^{-1}}}

\newcommand{\Heq}{\hat {H}_{\alpha}}
\newcommand{\HeqU}{\Heq}

\newcommand{\Fhu}{\overline F_{\alpha}}
\newcommand{\Chu}{\overline C_{\alpha}}
\newcommand{\Zhu}{\overline Z_{\alpha}}
\newcommand{\betaHU}{\overline \beta_{\alpha}}

\newcommand{\SMeq}{\stackon[-4pt]{SM}{\vstretch{1}{\hstretch{0.4}{\widehat{\phantom{\;\;\;\;\;\;\;\;}}}}}_{\alpha,q}}

\newcommand{\Csmu}{C_{\alpha,q}}
\newcommand{\betaSML}{\beta^{\, L}_{\alpha,q}}
\newcommand{\betaSME}{\beta^{\, E}_{\alpha,q}}

\newcommand{\Cseu}{C_{\alpha,r}}
\newcommand{\betaSEL}{\beta^{\, L}_{\alpha,r}}
\newcommand{\betaSEE}{\beta^{\, E}_{\alpha,r}}
\newcommand{\betaSE}{\beta_{\alpha,r}}

\newcommand{\betaSMU}{\beta_{\alpha,q}}

\newcommand{\sar}{s_{\alpha,r}}

\begin{document}

\begin{frontmatter}

\title{\Large \bf On the equivalence between four versions of thermostatistics\\ based on strongly pseudo-additive entropies}

\author[misasa]{Velimir M. Ili\'c\corref{cor}}
\ead{velimir.ilic@gmail.com}
\address[misasa]{Mathematical Institute of the Serbian Academy of Sciences and Arts, Kneza Mihaila 36, 11000 Beograd, Serbia}
\author[polidito]{Antonio Maria Scarfone}
\ead{antoniomaria.scarfone@cnr.it}
\address[polidito]{Istituto Sistemi Complessi (ISC--CNR) c/o Politecnico di Torino, Corso Duca degli Abruzzi 24, Torino, I-10129, Italy }
\author[eecibar]{Tatsuaki Wada}
\ead{tatsuaki.wada.to@vc.ibaraki.ac.jp}
\address[eecibar]{Region of Electrical and Electronic Systems Engineering, Ibaraki University, Nakanarusawa-cho, Hitachi, Ibaraki 316-8511, Japan}

\cortext[cor]{Corresponding author. E-mail:velimir.ilic@gmail.com, Tel.: +38118224492; fax: +38118533014.}

\begin{abstract}
The class of strongly pseudo-additive entropies, which can be
represented as an increasing continuous transformation of Shannon
and R\'enyi entropies, have intensively been studied in previous
decades. Although their mathematical structure has thoroughly been
explored and established by generalized Shannon-Khinchin axioms,
the analysis of their thermostatistical properties have mostly been
limited to special cases which belong to two parameter
Sharma-Mittal entropy class, such as Tsallis, Renyi and Gaussian
entropies.

In this paper we present a general analysis of the strongly
pseudo-additive entropies thermostatistics by taking into account
both linear and escort constraints on internal energy. We
develop two types of dualities between the thermostatistics formalisms. By the first one, the formalism of R\'enyi entropy is transformed in the formalism of SPA entropy under general energy constraint and, by the second one, the generalized thermostatistics which corresponds to the linear constraint is transformed into the one which corresponds to
the escort constraint. Thus, we establish the equivalence between four different thermostatistics formalisms based on R\'enyi and SPA entropies coupled with linear and escort constraints and we provide the transformation formulas.
In this way we obtain a general framework which is applicable to the wide class of entropies and constraints previously discussed in the
literature.

As an example, we rederive maximum entropy distributions for Sharma-Mittal entropy and we establish new relationships between the corresponding thermodynamic potentials. We obtain, as special cases, previously developed expressions for maximum entropy distributions and
thermodynamic quantities for Tsallis, R\'enyi and Gaussian entropies. In addition, the results are applied for derivation of thermostatistical relationships for supra-extensive entropy, which has not previously been considered.

{\em Keywords:} Strongly pseudo-additive entropy, Legendre
structure, generalized thermodynamics, maximum entropy, energy constraint,
Sharma-Mittal entropy, supra extensive entropy
\end{abstract}


\end{frontmatter}

\section{Introduction}
The class of strongly pseudo-additive (SPA) entropies is the most general
class of entropies which satisfies generalized Shannon-Khinchin
axioms \cite{Ilic_Stankovic_14}. SPA entropies can be represented as an
increasing continuous transformation of Shannon and R\'enyi
entropies and some well known generalized entropies, such as
Sharma-Mittal, Tsallis and Gaussian ones, belonging to this class.
In the past, the SPA entropies have widely been explored in the
context of statistics \cite{Salicru_et_al_93}, \cite{esteban1995summary}, computer sciences \cite{ciuperca2011computation} and quantum mechanics
\cite{Zozor_et_al_14}, and their ubiquitous mathematical structure has been recognized in the context of formal group theory where they are also known as Z-entropies \cite{tempesta2016formal}. A special importance of SPA entropies has been recognized in thermostatistics based on generalized maximum
entropy principle \cite{jaynes1957information}, \cite{jizba2019maximum}, which is the basis of non-extensive statistical mechanics \cite{Tsallis_et_al_98}, \cite{Lenzi_et_al_00}, \cite{Frank_Plastino_02}, \cite{Oikonomou_07}.

In accordance to the generalized maximum entropy principle, the equilibrium state of a system is determined by the configuration for which the
generalized entropy attains the maximum value, subject to
appropriately chosen mean energy constraint, which ensures that the
energy average of the system is fixed to a predefined value. The
average is usually defined as a linear expectation with respect to
the original distribution (referred as the first choice
constraint), or with respect to the escort distribution (the third
choice constraint) \cite{Tsallis_et_al_98}. Thus, R\'enyi entropy
is analyzed in \cite{Lenzi_et_al_00}, Tsallis entropy in
\cite{Tsallis_et_al_98}, Sharma-Mittal entropy in \cite{Frank_Plastino_02}
and Gaussian entropy in \cite{Oikonomou_07}, were the maximum entropy
distributions are derived and it is shown that the corresponding
thermodynamic quantities satisfy Legendre structure. On the
other hand, although the Legendre structure is satisfied in the
general case of entropies and constraints \cite{mendes1997some},
\cite{scarfone2016consistency}, \cite{yamano2000robust}, detailed
and general analysis of thermostatistics which is based on SPA
entropies still has not been conducted.

In this paper we consider generalized thermostatistics for SPA
entropies by taking into account linear and escort constraints. We
develop a generalized entropy duality principle by which the
thermostatistics formalism of R\'enyi entropy is transformed in
the thermostatistics formalism of SPA entropy, we establish the conditions for the transformation which preserve Legendre structure of thermostatistics and we provide the transformation formulas. In addition, we derive the linear to escort energy constraints duality for SPA entropy, by which the generalized thermostatistics which corresponds to the linear constraints can be transformed into the one which corresponds to the escort constraint. Thus, we establish the equivalence between four different thermostatistic formalisms based on R\'enyi and SPA entropies coupled with linear and escort constraints. In this way
we obtain a general framework which is applicable to the wide
class of entropies and constraints previously discussed in the
literature.

As an example, we rederive maximum entropy distributions for Sharma-Mittal entropy which is coupled with linear \cite{scarfone2006thermal}, \cite{lenzi2012extensive} and
escort constraints \cite{Frank_Plastino_02} and we establish new relationships between the corresponding thermodynamic quantities. We obtain, as special cases, previously developed expressions for maximum entropy distributions and
thermodynamic quantities for Tsallis entropy \cite{Tsallis_et_al_98}, \cite{Wada_02}, \cite{abe2001nonextensive}, R\'enyi entropy \cite{bashkirov2004maximum}, \cite{Lenzi_et_al_00} and Gaussian entropy \cite{Oikonomou_07}. In addition, the results are applied for derivation of thermostatistical relationships for supra-extensive entropy \cite{masi2005step},
which has not previously been considered.

The paper is organized as follows. In Section \ref{prel} we
present basic facts about SPA entropies, with a special attention to Sharma-Mittal and supra-extensive entropies. In Section \ref{gts} we introduce the generalized thermostatistical relationships in the case of generalized
entropies. These relationships are further used in Section
\ref{R-SPA dual} for the derivation of the R\'enyi-SPA duality.
The entropy constraint duality is considered in Section \ref{L_E
dual}, and the results are combined in Section \ref{me}, where we
present the equivalence between four different versions of
thermostatisitcs. In Section \ref{applic}, we derive the
thermostatistical relationships among Sharma-Mittal, supra-extensive and R\'enyi entropies. Concluding remarks are given in
Section \ref{conc}.

\section{Preliminiaries}
\label{prel}

\subsection{Generalized logarithm, exponential and
pseudo-addition}
\label{prel: log ex pa}

Let the sets of positive and nonnegative real
numbers be denoted with $\sR^+$ and $\sRp$, respectivelly, and
let $\h: \sR \ra \sR$ be an increasing continuous (hence
invertible) function such that $h(0)=0$. The generalized logarithm
$\Log: \sR_0 \ra \sR$ and its inverse, the generalized
exponential, are defined as:
\begin{equation}
\label{prel: Log(u)} \Log_h (u) \defeq h \lb \ln \ u \rb
\quad\text{and}\quad
\Exp(v) \defeq \Log^{-1}(v) = \text e^{h^{-1}(v)}.
\end{equation}
The pseudo-addition operation $\op$ is generated using $h$ as \cite{curado2016new}
\begin{equation}
\label{prel: h(x+y)=h(x) op h(y)} \h(x+y) = \h(x)
\op \h(y); \quad x, y \in \sR.
\end{equation}
which can be rewritten in terms of the generalized logarithm by
settings $x=\log u$ and $y =\log v$ so that
\begin{equation}
\label{prel: Log(u+v)=Log(u) op Log(v)} \Log(u \cdot v) = \Log(u)
\op \Log(v); \quad u, v \in \sR_+.
\end{equation}
An important type of the mapping $h$ is
$h_q: \sR \ra \sR$ defined in
\begin{equation}
\label{prel: h_q}
h_q(x) \defeq %
\begin{dcases}
\quad\quad x, &\mbox{ for }\quad q = 1 \\
\frac{e^{(1-q) \ x} - 1}{1-q}, \quad
&\mbox{ for }\quad q \neq 1
\end{dcases}
\end{equation}
and its inverse is given in
\begin{equation}
\h_q^{-1}(x) \defeq %
\begin{dcases}
\quad\quad x, &\mbox{ for }\quad q = 1 \\
\frac{1}{1-q} \ln((1-q) x +1), \quad
&\mbox{ for }\quad q \neq 1
\end{dcases}.
\end{equation}
with $q>0$. By the setting $h \defeq h_q$, the generalized
logarithm and exponential \eqref{prel: Log(u)} reduce to the
Tsallis $q$-logarithm, defined in
\begin{equation}
\Log_q \defeq %
\begin{dcases}
\quad\log x, &\mbox{ for }\quad q = 1 \\
\frac{x^{(1-q)} - 1}{1-q}, \quad
&\mbox{ for }\quad q \neq 1
\end{dcases},
\end{equation}
and $q$-exponential, defined in
\begin{equation}
\quad\quad
\Exp_q(y) \defeq%
\begin{dcases}
\quad \text e^y, &\mbox{ for }\quad q = 1 \\
\lb 1+(1-q) y \rb^{\frac{1}{1-q}}
&\mbox{ for }\quad q \neq 1.
\end{dcases}
\end{equation}
It holds that $\h_q(x+y)=\h_q(x) \op_q \h(y)$, where $\op_q$ is
$q$-addition \cite{tsallis1994numbers}, \cite{Nivanen_Wang__03}.

\subsection{Strongly pseudo-additive entropy}

Let the set of all $n$-dimensional distributions be denoted in
\begin{equation}
\label{prel: Delta_n}
   \Delta_n \defeq \left\{ (p_1, \dots , p_n) \Big\vert \; p_k \in \sRp,
      \sum_{k=1}^{n} p_k = 1 \right \};\quad  n>1.
\end{equation}
R\'enyi entropy is a function $\Ra: \Delta_n \ra \sR$ defined in
\begin{equation}
\label{prel: R_a}%
\Ra(\mcP) %
\defeq
\begin{dcases}
- \sum_{k=1}^{n} p_k \log
(p_k) \quad  &\mbox{for} \quad\quad \alpha =1\\
\frac{1}{1 - \alpha} \log \left( \sum_{k=1}^{n} p_k^{\alpha}
\right), \quad &\mbox{for} \quad\quad  \alpha \neq 1
\end{dcases}
\end{equation}
with $\alpha>0$.
Notably, it can be derived as the unique function that satisfies generalized Shannon-Khinchin (SK) axioms, which state that the entropy should be continuous, maximized for uniform distribution, expandable, and strongly additive. The latest property states that the entropy of a joint system can be represented as sum of the entropy of one system and the (generalized) conditional entropy of another, with respect to the first one \cite{Jizba_Arimitsu_04}.

An important generalization of R\'enyi entropies is the class of strongly pseudo-additive (SPA) entropies $\Hah$ which can be represented as the $h$ transformation of R\'enyi entropy \cite{Ilic_Stankovic_14}:
\begin{equation}
\label{prel: H=h(R)}%
\Hah(\mcP) \defeq  \h \lb \Ra(\mcP) \rb. %
\end{equation}
The class of SPA entropies can uniquely be derived from the generalized Shannon-Khinchin axioms if the strong additivity is replaced with the more general property, the strong pseudo-additivity. More specifically, let $\mcP = (p_1, \dots, p_n) \in \Delta_n$,
$PQ = (r_{11}, r_{12}, \dots, r_{nm}) \in \Delta_{nm}$, $n, m
\in \sN$, $n>1$ such that $p_i = \sum_{j=1}^m r_{i j}$, and
$Q_{ | i} = (q_{1|i}, \dots, q_{m|i}) \in \Delta_m$, where
$q_{j|i} = r_{ij}/p_i$ and $q \in \sR_+$ is some fixed parameter.
Then, the strong pseudo-additivity states that \cite{Ilic_Stankovic_14}
\begin{equation}
\label{ax 4} H_\alpha(PQ) = H_\alpha(P) \op H_\alpha(Q| P),
\end{equation}
where
\begin{equation}
H_\alpha(Q| P) = f^{-1} \left(\sum_{i=1}^n \escp{i} f(H_\alpha(Q_{ | i})) \right),
\end{equation}
$f$ is an invertible continuous function, $\alpha >
0$ is given parameter and
\emph{the $\alpha$-escort distribution}
$\mcP^{(\alpha)}\in \Delta_n$ of the distribution $\mcP \in
\Delta_n$ is given in
\begin{equation}
\label{L-E dual: P^(alpha)}%
\mcP^{(\alpha)}=(\escp{1},\dots,\escp{n}); \quad\quad
\escp{k}=\frac{p_k^\alpha}{\sum_{i=1}^n p_i^\alpha},\quad
k=1,\dots,n.
\end{equation}
Obviously, R\'enyi entropy is a special case of SPA entropies
which is obtained if $h\defeq\iden$, where $\iden$ stands for the
identity function. In the remaining part of the section we review some SPA entropies which were previously considered in statistical physics.

\subsubsection{Sharma-Mittal entropy}

An important special case of SPA entropies is the Sharma-Mittal
entropy \cite{Sharma_Mittal_75}, \cite{Frank_00} which is obtained
for $h \defeq h_q$, where $h_q$ is defined in \eqref{prel: h_q}.
Thus, the Sharma-Mittal entropy is given by
\begin{equation}
\label{prel: H_a,q=h_q(R_a)}%
\SMaq(P)\defeq h_q \lb \Ra(P)\rb =%
\frac{1}{1-q} \lb e^{(1-q)\Ra(P)} - 1\rb %
\end{equation}
and it reduces to the Shannon entropy \cite{Khinchin_57}, for
$q=\alpha=1$, to the Tsallis entropy, for $\alpha=q\neq 1$, to the
Gaussian entropy \cite{Frank_00}, for $q\neq1, \alpha=1$, and to
the R\'enyi entropy \cite{Renyi_07}, for $q=1, \alpha\neq 1$.
Note that in the case of the mapping $h_q$, it holds that
$\h_q(x+y)=\h_q(x) \op_q \h(y)$, so that Sharma-Mittal entropy is
strongly $q$-additive and it follows the $\op_q$ decomposition
rule \cite{Ilic_Stankovic_14}.

\subsubsection{Supra-extensive entropy}

Another important special case of SPA entropies is the
supra-extensive entropy \cite{masi2005step} which is obtained for
$h \defeq \sar$ where
\begin{equation}
\label{prel: s_ar} \sar(x)=%
h_\alpha \lb
h_r^{-1} \lb x \rb \rb =%
\frac{1}{1-\alpha} \lb \lb 1+(1-r)\ x \rb^{\frac{1-\alpha}{1-r}}
-1 \rb.
\end{equation}
Thus the supra-extnesive entropy has the form:
\begin{equation}
\label{prel: S_sup_ex} \SEag(P)\defeq \sar\lb \Ra(P)
\rb=\frac{1}{1-\alpha} \lb \lb (1-r)\Ra(P)+1
\rb^{\frac{1-\alpha}{1-r}} -1 \rb
\end{equation}
and reduces to R\'enyi entropy for $\alpha=r$ and to Tsallis
entropy for $r=1$.

\section{Legendre structure of SPA entropy
thermodynamics} \label{gts}

Let the energy levels of $n$-states system ($n\in \sN$) be
represented by the vector
$\varepsilon=(\varepsilon_1,\dots,\varepsilon_n)$ and let the
probability distribution on states be denoted with
$P=(p_1,\dots,p_n) \in \Delta_n$.
The equilibrium distribution $\hat P \in \Delta_n$ of a system is
determined by the maximum-entropy (ME) principle, as a state for
which the entropy $\Hah$ attains the maximum value, subject to the
internal energy constraint $U=u(P,E)$, where $u(P,E)$ is an appropriately
chosen regular function, which defines the mean value of $E$,
\begin{equation}
\label{gts: P=argmax H}
\hat P \defeq \argmax_{ P \in \Delta_n} \left\{ \Hah(P)\ \Big| \  u\lb P,E\rb=U \right\}.
\end{equation}
and the maximum entropy value is given by
\begin{equation}
\label{gts: H=H(P)}
\Heq \defeq \Hah(\hat P).
\end{equation}
Note that the equilibrium distribution \eqref{gts:
P=argmax H} and the maximum entropy \eqref{gts: H=H(P)} depend on
the choice of the constraint function $u(P,E)$ as well as on internal energy $U$. However, in the following paragraphs we implicitly assume that the function $u(P,E)$ is fixed, so that we will consider $\hat P$ and $\Heq$
as functions of $U$ only and we study the generalized themrostatistics based on the equilibrium entropy \eqref{gts: H=H(P)}.

In this paper, we limit our discussion to the systems for which the generalized beta parameter (coldness), defined in
\begin{equation}
\label{legstr: b=dH/dU}
\betaHU\defeq\frac{\pd \HeqU}{\pd U}, 
\end{equation}
is a nonzero,
continuously differentiable monotonic function of $U$, with the continuous first derivative for all $\alpha>0$, and we denote the conditions with LSH, i.e.
\begin{equation}
\label{legstr: b, db/dU neq 0}
\text{LSH:}\quad
\quad \betaHU \in C^1, \quad \betaHU \neq 0 \quad \text{and} \quad \frac{\pd \betaHU}{\pd U}\neq 0,
\end{equation}
where $C^1$ denotes the space of continuously differentiable functions.
Due to the monotonicity of $\betaHU$, $U$ can be uniquely represented as a function of $\betaHU$. Therefore, the generalized entropy \eqref{gts: H=H(P)} which is a function of $U$, can be represented in terms of
$\betaHU$ by the substitution $U\defeq U(\betaHU)$. With this
settings, we can define generalized log partition function as
Legendre transform of generalized entropy, like in
the Boltzman-Gibbs thermostatistics:
\begin{equation}
\label{legstr: log Z=H-b*U}
\ln \Zhu \defeq
\HeqU - \betaHU U
\end{equation}
In addition, standard relationships also holds in this case
\begin{equation}
\label{legstr: dH/db & U=dZ/db}
\frac{\pd \HeqU}{\pd \betaHU}=%
\betaHU \frac{\pd U}{\pd \betaHU} \quad%
\quad\text{and}\quad
U=-\frac{\pd \ln \Zhu}{\pd\ \betaHU},%
\end{equation}
and generalized free energy can be defined (as a function of
$\betaHU$) with:
\begin{equation}
\label{legstr: F=U-T*H}
\Fhu \defeq U - \frac{1}{\betaHU} \HeqU = -\frac{1}{\betaHU}\ln \Zhu,
\end{equation}
where the last equality follows from \eqref{legstr: log Z=H-b*U}.
Note that, using \eqref{legstr: dH/db & U=dZ/db}, we can derive
the standard inverse relationship:
\begin{equation}
\label{legstr: H=dF/dT}
\HeqU=\betaHU^2 \frac{\pd \Fhu}{\pd\ \betaHU}.%
\end{equation}
Moreover, the generalized specific heat capacity can be defined as
\begin{equation}
\label{legstr: C=dU/dT}
\Chu \defeq
-\betaHU\frac{\pd \HeqU}{\pd \betaHU}=%
-\betaHU^2\frac{\pd U}{\pd \betaHU},
\end{equation}
where the righthand side equalities follow from \eqref{legstr: dH/db & U=dZ/db}.
The equations \eqref{legstr: b=dH/dU}-\eqref{legstr: H=dF/dT}
describe the Legendre structure of thermodynamics which is based on
SPA entropies and is valid for any $h$ and any $u$. In the
following sections we draw the relationships between the thermodynamics which are derived for $h \defeq 1$ (R\'enyi entropy) and an arbitrary $h$ (SPA entropy), for two most commonly used types of constraints (linear and escort ones).

\begin{remark}\rm
Due to the continuity of $\betaHU$ and its derivative
per $U$, the LSH conditions \eqref{legstr: b, db/dU neq 0} can be
rewritten as \begin{equation}
\betaHU \neq 0 \quad \text{and}\quad \Chu \neq 0,
\end{equation}
with $\Chu$ being continuous. The LSH conditions ensure that the entropy is monotonic and strictly convex or concave function of internal energy which implies the uniqueness of its Legendre transform in \eqref{legstr: log Z=H-b*U}. From physical point of view, the conditions should be more restrictive to ensure some desirable properties such as thermodynamic stability \cite{Ramshaw_95}, \cite{scarfone2006thermal}. However, the general discussion from the paper will also hold with these more restrictive conditions.
\end{remark}

\section{Duality between R\'enyi and SPA entropies
(R-SPA duality)}\label{R-SPA dual}

Like in the previous section, we consider $n$-states system ($n \in N$) which is represented by the vector $\varepsilon=(\varepsilon_1,\dots,\varepsilon_n)$ and optimization procedure \eqref{gts: P=argmax H} under the general constraint $u\lb P,E\rb=U$, this time for R\'enyi entropy (the case $h =\iden$). Thus, we consider the optimal distribution:
\begin{equation}
\label{R-SPA dual: P=argmax R}%
\Pru =\argmax_{ P \in \Delta_n} \left\{ \Ra(P)\ \Big| \  u\lb P,E\rb=U \right\}
\end{equation}
for which we indicate the maximum entropy
\begin{equation}
\label{R-SPA dual: R=R(P)=log Z} \ReqU = \Ra(\Pru).
\end{equation}
Note that the optimal distribution in the case of SPA entropies
\begin{equation}
\label{R-SPA dual: P=argmax H} \Pru
= \argmax_{P \in \Delta_n} \left\{ \Hah(P)\ \Big| \  u\lb P,E\rb=U 
 \right\}
\end{equation}
has the same expression as obtained in \eqref{R-SPA dual: P=argmax
R} since $\Hah = h(\Ra)$ and $h$ is increasing, so that the general relationship between SPA and R\'enyi entropy \eqref{prel: H_a,q=h_q(R_a)} also holds in the case of equilibrium and the maximum SPA entropy $\HeqU = \Hah(\Pru)$ can be expressed in terms of the maximum R\'enyi entropy \eqref{R-SPA dual: R=R(P)=log Z} and vice verse as
\begin{equation}
\label{R-SPA dual: H=h(R)} \HeqU = h(\ReqU)
\quad \Leftrightarrow \quad \ReqU = h^{-1}(\HeqU).
\end{equation}
Following the discussion in Section \ref{gts} (by the setting $h = \iden$), we are able to define all thermodynamic quantities that corresponds to R\'enyi entropy and, for notational convenience, we omit the bars and denote them as $\betaRU$, $\Tru$, $\Fru$ and  $\Cru$.
In order to ensure the Legendre structure of R\'enyi thermodynamics we also assume LSH conditions, which are in the case of R\'eni entropy denoted with LSR and are given in%
\begin{equation}
\label{R-SPA dual: b, db/dU neq 0}
\text{LSR:} \quad \betaRU \in C^1, \quad%
\betaRU \neq 0 \quad \text{and}
\quad \frac{\pd \betaRU}{\pd U}\neq 0.
\end{equation}
Two questions that naturally arises at this point are:
\begin{enumerate}

\item What are the choices of function $h$ for which the Legendre structure of thermodynamics preserved under the transformation from the R\'enyi to the SPA formalism and vice verse?

\item What are the transformation formulas for thermodynamic potentials from the R\'enyi to the SPA formalism and vice verse?

\end{enumerate}
To answer the questions we first derive additional conditions under which the LSH conditions are satisfied, if LSR conditions are assumed. We relate the SPA coldness $\betaHU$ and R\'enyi coldness $\betaRU$ by taking the partial derivative of \eqref{R-SPA dual: H=h(R)} over $U$ and using the chain rule:
\begin{equation}
\label{R-SPA dual: b_h=h'*b_r & t_r=h'*t_h}
\betaHU=h^\prime(\ReqU)\ \betaRU.
\end{equation}
Obviously, the LSH condition, $\betaHU\neq 0$ directly follows from $\betaRU\neq0$ since the function $h$ is increasing. However, the second one, i.e. the monotonicity of $\betaHU$,
\begin{equation}
\label{R-SPA dual: db_h/dU }
\frac{\pd \betaHU}{\pd U}=%
\frac{\pd \betaHU}{\pd \betaRU}\frac{\pd \betaRU}{\pd U} \neq 0,
\end{equation}
is satisfied if and only if $\pd \betaHU/\pd \betaRU$ is continuous and non-zero,
\begin{align}
\label{R-SPA dual: db_h/db_r}
\frac{\pd \betaHU}{\pd \betaRU}&=%
\h^\dprime(\ReqU)\ \betaRU\  \frac{\pd \ReqU}{\pd \betaRU}
+\h^\prime(\ReqU)
=\h^\prime(\ReqU) - \h^\dprime(\ReqU)\ \Cru
\neq 0,
\end{align}
where the R\'enyi heat capacity is expressed using
the relationship \eqref{legstr: C=dU/dT},
\begin{equation}
\label{R-SPA dual: C=-b*dR/db} \Cru=-\betaRU\  \frac{\pd
\ReqU}{\pd \betaRU}.
\end{equation}

Therefore, if LSR conditions are assumed, LSH conditions also hold, providing that the additional condition (denoted with HC) is satisfied:
\begin{equation}
\label{R-SPA dual: Cr neq h'/h"}
\text{HC:}\quad \h^\prime(\ReqU) \neq \h^\dprime(\ReqU)\ \Cru,
\end{equation}
which can be denoted in
\begin{equation}
\text{LSR}\quad + \quad \text{HC} \quad \Rightarrow \quad \text{LSH}.
\end{equation}
In addition, since $\pd \betaHU/\pd \betaRU$ and $\pd \betaRU/\pd \betaHU$ are continuous and non-zero simultaneously, the opposite direction can also be derived,
\begin{equation}
\text{LSH}\quad + \quad \text{HC} \quad \Rightarrow \quad \text{LSR}.
\end{equation}
In other words, if HC conditions are assumed, LSH and LSR conditions are equivalent and Legendre structure of the R\'enyi and SPA formalisms is simultaneously preserved.

\begin{remark}\rm
In general, the satisfiability of the HC condition depends on the choice of the function $h$, value of R\'enyi entropy $\Req$ and the heat capacity $\Cru$ and, consequently, on the choice of $\varepsilon$ and $\alpha$, as well as on the choice of the internal energy  constraint $u(P,E)$. Therefore, in particular cases, it is possible to specify further the properties of $h$ that ensure the inequality \eqref{R-SPA dual: Cr neq h'/h"} and the equivalence of LSH and LSR conditions. As examples, we consider the following cases.

\begin{enumerate}

\item In the case of positive R\'enyi heat capacity, which usually appears as the requirement for thermodynamic stability \cite{Ramshaw_95}, \cite{Wada_02}, Legendre structures of the R\'enyi and the SPA entropy $h(\Req)$ formalisms are preserved simultaneously for any concave function $h$ (concavity of $h$ is a sufficient condition).
\item In the case of $\Cru = 1$, which appears in 1-dimensional classical harmonic oscillator with two particles \cite{mendes1997some}, concavity of $h$ is too restrictive. Here, the HC condition has the simple form $h^\prime(x) \neq h^{\prime\prime}(x)$. By taking into account the definition of $h$ from the section \ref{prel: log ex pa}, we can easily obtain that HC requirement is satisfied for any choice of increasing continuous function $h$ such that $h(0) = 0$, which differs from $c\cdot h_0$ ($c>0$ and $h_0$ is defined by the setting $q=0$ in \eqref{prel: h_q}). Note that the case of $h=c\cdot h_0$ corresponds to (the multiple of) Sharma-Mittal entropy \eqref{prel: H_a,q=h_q(R_a)}, $\SMao$.
\item Oppositely to the case 1, if the R\'enyi heat capacity is negative, which appears in systems with long-range correlations \cite{dauxois2002dynamics}, \cite{Ramshaw_95}, Legendre structures is preserved simultaneously for R\'enyi and SPA thermostatistics for any convex function $h$ (convexity of $h$ is a sufficient condition).

\end{enumerate}

\end{remark}
In the following discussions, we assume that the HC condition is satisfied, so that LSR and LSH are equivalent and we derive the transformation formulas from one to another formalism.

The R\'enyi and SPA log partition functions are related using the equation \eqref{legstr: log Z=H-b*U}, once for the SPA, and once for the R\'enyi entropy, and by eliminating $U$ from the equations, so that
\begin{equation}
\frac{\HeqU - \ln \Zhu}{\betaHU} = \frac{\Req - \ln Z_\alpha}{\betaRU}.
\end{equation}
Thus, we obtain
\begin{equation}
\ln \Zhu = h(\Req) - h^\prime(\Req) (\Req - \ln Z_\alpha),
\end{equation}
and, vice verse,
\begin{equation}
\ln Z_\alpha = g(\HeqU) - g^\prime(\HeqU) (\HeqU - \ln \Zhu),
\end{equation}
where we set up $g=h^{-1}$ and used \eqref{R-SPA dual: H=h(R)} and \eqref{R-SPA dual: b_h=h'*b_r & t_r=h'*t_h}. The inverse relationships for the free energy can be straighforwardly computed in the same manner, so that
\begin{equation}
\Fhu = -\frac{1}{\betaRU} \frac{h(\Req)}{h^\prime(\Req)}  +\frac{\Req}{\betaRU} +\Fru
\end{equation}
and, vice verse,
\begin{equation}
\Fru = -\frac{1}{\betaHU} \frac{g(\HeqU)}{g^\prime(\HeqU)}  +\frac{\HeqU}{\betaHU} +\Fhu.
\end{equation}

In addition, by usage of chain rule and inverse
function derivative formula, and taking into account the equations
\eqref{R-SPA dual: H=h(R)}, \eqref{R-SPA dual: b_h=h'*b_r &
t_r=h'*t_h}, \eqref{R-SPA dual: db_h/db_r} and \eqref{R-SPA dual:
C=-b*dR/db}, we obtain the connection between the corresponding
specific heats:
\begin{align}
\label{R-SPA dual: C=-b*dH/db}
\Chu&= -\betaHU\frac{\pd \HeqU}{\pd \betaHU}=%
\dfrac{-\betaHU\dfrac{\pd \HeqU}{\pd \ReqU}\dfrac{\pd \ReqU}{\pd \betaRU}}%
{\dfrac{\pd \betaHU}{\pd \betaRU}}=
\frac{\h^\prime(\ReqU)^2 \Cru}%
{\h^\prime(\ReqU) - \h^\dprime(\ReqU)\ \Cru},
\end{align}
or equivalently
\begin{equation}
\label{R-SPA dual: 1/Cr = h"/h' + h'*1/Cr}
(\Cru)^{-1}
= \frac{\h^\dprime(\ReqU)}{\h^\prime(\ReqU)}+
\h^\prime(\ReqU)
(\Chu)^{-1}.
\end{equation}
Note that the HC condition \eqref{R-SPA dual: Cr neq h'/h"} ensures that $\Cru$ and $\Chu$ are finite and non-zero simultaneously.

\section{Entropy-constraints duality for SPA entropies (E-C duality)}\label{L_E dual}

The discussion so far is general and applicable to any type of
constraint $U=u(P,E)$, where $u(P,E)$ is an appropriately chosen
regular function. In this section, we specify the discussion to
two most important cases which are considered in statical
mechanics \cite{Tsallis_et_al_98}:
\begin{enumerate}
\item
Linear expectation, also referred as the first choice of the constraint:
$u(P,E)=\sum_{i=1}^n p_i E_i$, which generate the maximum
entropy distribution
\begin{align}
\label{L-E dual: P^L=argmax H}
\PrL &=%
\lb  \prL{1}, \dots, \prL{n} \rb\nonumber\\ &= \argmax_{P \in \Delta_n} \left\{ \Hah(P)\ \Big| \ \sum_i p_i
E_i=U
\right\}
\end{align}
and the equilibrium entropy
\begin{equation}
\label{L-E dual: H^L=H(P^L)} \HeqL=\Hah\lb \hat P_\alpha^L \rb.
\end{equation}

\item
{Escort expectation}, also referred as the third choice of the constraint: $u(P,E)=\sum_{i=1}^n P_i^{(\alpha)} E_i$, where the $\alpha$-escort distribution is defined in \eqref{L-E dual: P^(alpha)}.
The maximum entropy distribution for the $\alpha$-escort constraint is given in
\begin{align}
\label{L-E dual: P^E=argmax H}%
\PrE&=%
\lb  \prE{1}, \dots, \prE{n} \rb\nonumber\\
&=\argmax_{P \in \Delta_n} \left\{ \Hah(P)\ \Big| \ \sum_i
P_i^{(\alpha)}
E_i=U
\right\}
\end{align}
and the equilibrium entropy is
\begin{equation}
\label{L-E dual: H^E=H(P^E)} \HeqE=\Hah\lb \hat P_\alpha^E \rb.
\end{equation}

\end{enumerate}

The relationships among the quantities defined using linear and
escort constraints can be obtained using entropy-constraint (E-C)
duality principle, which can be stated as follows. From equation
\eqref{L-E dual: P^(alpha)} it is easy to find the inverse
relationship between a distribution $P \in \Delta_n$ and its
excort distribution $P^{(\alpha)} \in \Delta_n$, which is given in
\begin{equation}
\label{P=P^(alpha)^(alpha^(-1))}
P=\lb P^{(\alpha)} \rb^{(\alpha^{-1})},
\end{equation}
so that there exists one to one correspondence between two of
them, and the optimal escort distribution can be obtained if
$\PrEEsc$ is searched for in \eqref{L-E dual: P^E=argmax H},
instead of $\PrE$. In addition, it is straightforward to show that
\begin{equation}
\label{L-E dual: HahInv(P^(alpha))}
\HahInv\lb P^{(\alpha)} \rb=%
\Hah\lb P \rb\quad%
\Leftrightarrow\quad%
\Hah\lb P^{(\alpha^{-1})} \rb=%
\HahInv\lb P \rb
\end{equation}
with $\alpha>0$ and the equation \eqref{L-E dual: P^E=argmax H} can be converted into the equivalent one:
\begin{align}
\label{L-E dual: <PE_a>(a)=argmax H}%
\PrEEsc&=
\argmax_{P^{(\alpha)} \in \Delta_n} \left\{ \Hah(P)\ \Big| \
\sum_i P_i^{(\alpha)}
E_i=U 
\right\}\nonumber\\ &=%
\argmax_{P^{(\alpha)} \in \Delta_n} \left\{ \HahInv(P^{(\alpha)})\
\Big| \ \sum_i P_i^{(\alpha)}
E_i=U 
\right\}.
\end{align}
After formal substitution $P^{(\alpha)} \rightarrow P$ into second
equality of \eqref{L-E dual: <PE_a>(a)=argmax H}, we obtain the
form \eqref{L-E dual: P^L=argmax H} for ME problem of $\HahInv$
with the first choice constraint which has the solution $\PrLinv$.
Therefore, we obtain $\PrLinv=\PrEEsc$, or by taking into account
the relationship \eqref{P=P^(alpha)^(alpha^(-1))},
\begin{equation}
\label{L-E dual: <PE_a>(a)=PL_1/a esc}
\PrE= \laux\PrLinv \raux^{(\alpha^{-1})}.
\end{equation}
Finally if \eqref{L-E dual: <PE_a>(a)=PL_1/a esc} is combined with
\eqref{L-E dual: HahInv(P^(alpha))} we obtain the basic equation
of
E-C duality $\Hah\lb \PrE \rb=\HahInv\lb \PrL \rb
$, i.e. $\HeqE =\HeqLinv$. Note that the E-C duality hold for any
$\alpha>0$ and any choice of $h$ (including R\'enyi case
$h=\iden$), providing that the conditions \eqref{legstr: b, db/dU
neq 0} are satisfied. On the other hand, if the conditions are
satisfied for $\HeqE$, they are also satisfied for $\HeqL$ so
that the Legendre structure is preserved in both the cases.
Therefore, the dualities between all other thermostatistical
quantities also hold, so that we get the complete list of the E-C
duality relationships:
\begin{align}
\label{L-E dual:
TE_a,CE_a,FE_a,bE_a,ZE_a=TL_1/a,CL_1/a,FL_1/a,bL_1/a,ZL_1/a}
&\PrE= \laux\PrLinv \raux^{(\alpha^{-1})}, \quad
\ReqE=\ReqLinv, \quad
\HeqE=\HeqLinv,\nonumber\\  &\Che=
\ChlInv,\quad \Fhe=\FhlInv,\quad
\betaHE=\betaHLAm, 
\quad \Zhe=\ZhlInv.
\end{align}

\section{On the equivalence between four versions
of generalized thermostatistics}\label{me}

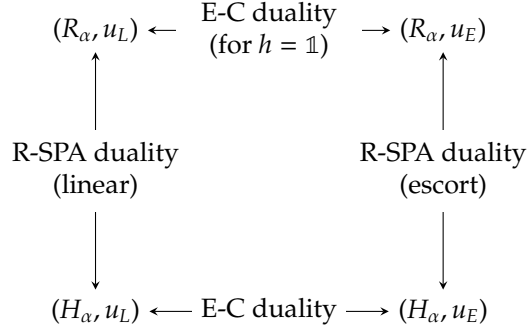
\begin{figure}[t]
\centering
\begin{tikzpicture}[descr/.style={fill=white}]
  \matrix (m) [matrix of math nodes,row sep=9em,column sep=9em,minimum width=2em]
  {
     (\Ra, u_L) &  (\Ra, u_E)\\
     (\Hah, u_L) & (\Hah, u_E)\\
     };
  \path[stealth-stealth]
    (m-1-1)  edge node [descr] {\begin{tabular}{c}E-C duality \\ (for $h\defeq \iden$)\end{tabular}} (m-1-2)
             edge node [descr] {\begin{tabular}{c}R-SPA duality\\ (linear)\end{tabular}} (m-2-1)
    (m-2-1.east|-m-2-2) edge node [descr] {E-C duality}(m-2-2)
    (m-1-2) edge node [descr] {\begin{tabular}{c}R-SPA duality\\ (escort)\end{tabular}} (m-2-2)
            ;
\end{tikzpicture}
\caption{Equivalence between R\'enyi and SPA thermostatistics under
linear and escort energy constraints} \label{fig: com diag}
\end{figure}

If SPA and R\'enyi entropies are combined with two different types
of constrains, four different thermostatistics (TS) can be
defined:
\begin{itemize}
    \item $(R_\alpha, u_L){{-TS}}$, which is based on R\'enyi entropy with linear constraints,
    \item $(R_\alpha, u_E){{-TS}}$, which is based on R\'enyi entropy with escort constraints,
    \item $(\Hah, u_L){{-TS}}$, which is based on SPA entropy with linear constraints, and
    \item $(\Hah, u_E){{-TS}}$, which is based on SPA entropy with escort
    constraints.
\end{itemize}
The formalisms are related by R-SPA duality derived in Section
\ref{R-SPA dual} and E-C duality derived in Section \ref{L_E
dual}. Thus, if we have computed the thermostatistical quantities
for $(R_\alpha, u_L)$ thermostatistics, we are able to compute the
quantities for $(\Hah, u_L)$ thermostatistics using the R-SPA duality, and the
quantities for $(R_\alpha, u_E)$ thermostatistics using the E-C
duality. Similarly, $(\Hah, u_E)$ thermostatistics can be derived
from either $(\Hah, u_L)$ thermostatistics or $(R_\alpha, u_E)$
thermostatistics. Note that all of the transformations derived in
Sections \ref{R-SPA dual} and \ref{L_E dual} are invertible, so
that the derivation path also exists from $(\Hah, u_E)$ to
$(R_\alpha, u_L)$ thermsotatistics.
Thus, {{starting from any of these thermostatistics we can derive
anothers}}, so that the formalisms are equivalent as
represented in Figure \ref{fig: com diag}.

The discussion will be illustrated by considering the maximum
entropy distribution $(R_\alpha, u_L)$ thermostatistics which can
be derived using Lagrangian optimization
\cite{bashkirov2004maximum}:
\begin{equation}
\label{me: pRL}%
\prL{i}= e^{-\ReqL} 
\left[1 - \frac{\alpha-1}{\alpha}\ \betaRL \DeltaE
\right]_+^{\frac{{1}}{\alpha-1}}, %
\end{equation}
where
\begin{equation}
\DeltaE=E_i-U
\end{equation}
for $i=1,\dots,n$.
As we noted before, $\Hah$ and $\Ra$ are optimized by the same
maximum entropy distribution \eqref{R-SPA dual: P=argmax R} algthough the generalized coldness $\betaHL$ depends on $h$ and is
related to $\betaRL$ in accordance to equation \eqref{R-SPA dual:
b_h=h'*b_r & t_r=h'*t_h}. Thus, ME distribution can be represented
in the form
\begin{equation}
\label{me: pHL}%
\prL{i}= e^{-\ReqL} \left[1 - \frac{\alpha-1}{\alpha}\
\frac{\betaHL}{h^\prime(\ReqL)} \DeltaE
\right]_+^{\frac{1}{\alpha-1}},
\end{equation}
for $i=1,\dots,n$, which follows left-hand line of Figure \ref{fig: com diag} and
corresponds to $(\Hah, u_L)-TS$.

The optimal distribution which corresponds to $(R_\alpha, u_E)$
thermostatistics can be derived following the upper line of the
Figure \ref{fig: com diag}. From the equalities
\eqref{L-E dual:
TE_a,CE_a,FE_a,bE_a,ZE_a=TL_1/a,CL_1/a,FL_1/a,bL_1/a,ZL_1/a},
we obtain
\begin{align}
\label{me: pRLInv}%
\prLInv{i}&= e^{-\ReqLinv} 
\left[1 - \frac{\alpha^{-1}-1}{\alpha^{-1}}\ \betaRLInv \DeltaE
\right]_+^{\frac{1}{\alpha^{-1}-1}}\nonumber\\&=%
e^{-\ReqE} 
\left[1 - (1-\alpha)\ \betaRE \DeltaE
\right]_+^{\frac{\alpha}{1-\alpha}}
\end{align}
Since $\sum_i \prLInv{i}^{\alpha^{-1}} = e^{(1-\alpha^{-1})
\ReqLinv}=e^{(1-\alpha^{-1}) \ReqE}$, the equality \eqref{L-E
dual: <PE_a>(a)=PL_1/a esc},
can be written as:
\begin{equation}
\label{me: pRE}%
\prE{i}= %
\frac{\prLInv{i}^{\alpha^{-1}}}{\sum_i \prLInv{i}^{\alpha^{-1}}}=
e^{-\ReqE} \left[1 - (1-\alpha)\ \betaRE \DeltaE
\right]_+^{\frac{1}{1-\alpha}}, 
\end{equation}
for $i=1,\dots,n$. Finally, the optimal distribution which corresponds to $(\Hah,
u_E)$ thermostatistics can be derived in similar manner, following
lower or right-hand line in the Figure \ref{fig: com diag}, so
that we obtain:
\begin{equation}
\label{me: pHE}%
\prE{i}= e^{-\ReqE} \left[1 - (1-\alpha)\
\frac{\betaHE}{h^\prime(\ReqE)} \DeltaE
\right]_+^{\frac{1}{1-\alpha}},
\end{equation}
for $i=1,\dots,n$.

\section{Applications} \label{applic} In this
section, we illustrate the previous discussion for Sharma-Mittal
and supra-extensive thermostatistics and derive their relationships to R\'enyi one.

\subsection{Generalized thermostatistics for Sharma-Mitttal entropy}

Sharma-Mittal entropy $\SMeq \defeq h_q\lb \ReqU \rb$ is obtained from the equation \eqref{prel: H=h(R)} by the setting $h\defeq h_q$, where $h_q$ is given in \eqref{prel: h_q}.
In this case we can compute
\begin{equation}
\label{prel: h_q'} h_q^\prime(\Req) = e^{(1-q) \ReqU} =
1 + (1-q) \SMeq.
\end{equation}
As mentioned before, the generalized coldness $\betaHU$ depends on the parameter $\alpha$ and on the mapping $h$. In the case of the mapping $h_q$, the dependence is translated into dependence on the pair of the parameters $\alpha$ and $q$. Thus, Sharma-Mittal coldness will be denoted as $\betaSMU$ and can be related to the R\'enyi coldness $\betaRU$, for an arbitrary energy constraint $u\lb P,E\rb=U$, using the equations
\eqref{R-SPA dual: b_h=h'*b_r & t_r=h'*t_h}:
\begin{equation}
\label{R-SPA dual: 1/Tr=(1+(1-q)*H)*1/Tsm}
\betaSMU =\lb 1 + (1-q) \SMeq \rb \betaRU.
\end{equation}
In addition, we have
\begin{equation}
\label{prel: h_q''} h_q^{\dprime}(\Req)
=(1-q)\ e^{(1-q)\ \ReqU}=(1-q) \lb 1 + (1-q) \SMeq \rb
\end{equation}
with
\begin{equation}
\frac{h_q^\prime(\Req)}{h_q^\dprime(\Req)} = \frac{1}{1-q},
\end{equation}
so that the HC condition \eqref{R-SPA dual: Cr neq h'/h"}, which preserve Legendre structure of Sharma-Mittal thermostatistics, has the form:
\[
\Cru \neq \frac{1}{1-q}.
\]
We can relate Sharma-Mittal heat capacity $\Csmu$ with R\'enti
heat capacity $\Cru$ using the equation \eqref{R-SPA
dual: 1/Cr = h"/h' + h'*1/Cr}:
\begin{align}
\label{R-SPA dual: 1/Cr of 1/Csm}%
\laux\Cru\raux^{-1}&=%
(1-q)+ \lb 1 + (1-q) \SMeq \rb \laux\Csmu\raux^{-1}\nonumber\\ &=
(1-q)+ e^{(1-q)\Req}  \laux\Csmu\raux^{-1}.
\end{align}
Thus, we have derived new relationships which reduce to previously
derived special cases of Tsallis entropy
\cite{abe2001nonextensive} (for $\alpha=q$) and Gausisan entropy
\cite{Oikonomou_07} (for $\alpha=1$).

The maximum entropy distributions which correspond to
Sharma-Mittal entropy can be obtained by the substitution of the
expression \eqref{prel: h_q'} in the expressions \eqref{me: pHL}
and \eqref{me: pHE}. Therefore, for the first choice
theromstatistics we rederive the ME forms from \cite{scarfone2006thermal} and \cite{lenzi2012extensive}:
\begin{equation}
\label{me: pSML}%
\prL{i}=%
e^{-\Req^{L}} \left[1 - \frac{\alpha-1}{\alpha}\
\frac{\betaSML}{e^{(1-q) \ReqL}} \DeltaE
\right]_+^{\frac{1}{\alpha-1}},%
\end{equation}
for $i=1,\dots,n$, and for the third choice theromstatistics we obtain
the ME form derived \cite{Frank_Plastino_02}:
\begin{equation}
\label{me: pSME}%
\prE{i}= e^{-\ReqE} \left[1 - (1-\alpha)\ \frac{\betaSME}{e^{(1-q)
\ReqE}} \DeltaE \right]_+^{\frac{1}{1-\alpha}},%
\end{equation}
for $i=1,\dots,n$. In addition, by special choices of the parameters, the ME
distributions \eqref{me: pSML} and \eqref{me: pSME} reduces to the
previously derived ME distributions for Shannon entropy ($\alpha=q=1$), R\'enyi entropy ($q=1$) \cite{Lenzi_et_al_00}, Tsallis entropy ($\alpha=q$)
\cite{Tsallis_et_al_98} and Gaussian entropy ($\alpha=1$)
\cite{Oikonomou_07}.

\subsection{Generalized thermostatistics for supra-extensive entropy}

The presented framework can also be used for the
derivation of thermostatistical relationships for the
supra-extensive entropy $\SEeq \defeq h_q\lb \ReqU \rb$ introduced
in \cite{masi2005step}, which is obtained if the SPA entropy
\eqref{prel: H=h(R)} is defined for $h\defeq \sar$, where $\sar$ is given in \eqref{prel: s_ar}. For the sake of simplicity, we will keep the same notation for generalized coldness and heat capacity as in the previous section.
The first derivative of $\sar$ can be expressed as
\begin{equation}
\label{prel: s_qa'}
\sar^\prime(\ReqU)=%
\lb (1-r)\ \ReqU+1 \rb^{\frac{1-\alpha}{1-r}-1}
\end{equation}
so that the relationships among the supra-extensive
coldness $\betaSMU$ and the R\'enyi coldness $\betaRU$, for an
arbitrary energy constraint $U=u(P,E)$, can be derived using the
equations \eqref{R-SPA dual: b_h=h'*b_r & t_r=h'*t_h}:
\begin{equation}
\label{R-SPA dual: 1/Tr=(1+(1-q)*H)*1/Tse}
\betaSE =
\lb (1-r)\ \ReqU+1 \rb^{\frac{1-\alpha}{1-r}-1} \betaRU.
\end{equation}
In addition, the second derivative of $\sar$ can be expressed as
\begin{equation}
\label{prel: s_qa''}
\sar^\dprime(\ReqU)=%
(r-\alpha)\lb (1-r)\ \ReqU+1 \rb^{\frac{1-\alpha}{1-r}-2}
\end{equation}
with
\begin{equation}
\frac{\sar^\prime(\ReqU)}{\sar^\dprime(\ReqU)} =%
\frac{(1-r)\ \ReqU +1}{r-\alpha},
\end{equation}
so that the HC condition \eqref{R-SPA dual: Cr neq h'/h"}, which preserve Legendre structure of Sharma-Mittal thermostatistics, has the form:
\[
\Cru \neq \frac{(1-r)\ \ReqU +1}{r-\alpha}.
\]
The supra-extensive heat capacity $\Cseu$ and
the R\'enyi heat capacity $\Cru$ are related using equation \eqref{R-SPA dual:
1/Cr = h"/h' + h'*1/Cr}:
\begin{equation}
(\Cru)^{-1} =
\frac{r-\alpha}{(1-r)\ \ReqU +1}+%
\lb (1-r)\ \ReqU+1 \rb^{\frac{1-\alpha}{1-r}-1}(\Cseu)^{-1}.
\end{equation}
The maximum entropy distributions which correspond to
supra-extensive entropy can be obtained by the substitution of the
expressions \eqref{prel: s_qa'} in the expressions \eqref{me: pHL}
and \eqref{me: pHE}. Thus, for the first choice theromstatistics
we obtain
\begin{equation}
\label{me: pSEL}%
\prL{i}= e^{-\Req^{L}} \left[1 - \frac{\alpha-1}{\alpha}\
\lb (1-r)\ \ReqL+1 \rb^{1-\frac{1-\alpha}{1-r}}
\betaSEL
\DeltaE \right]_+^{\frac{1}{\alpha-1}},
\end{equation}
for $i=1,\dots,n$, and for the third choice theromstatistics we obtain
\begin{equation}
\label{me: pSEE}%
\prE{i}= e^{-\ReqE} \left[1 - (1-\alpha)\
\lb (1-r)\ \ReqE+1 \rb^{1-\frac{1-\alpha}{1-r}} \betaSEE \DeltaE
\right]_+^{\frac{1}{1-\alpha}}, 
\end{equation}
for $i=1,\dots,n$. Note that all the expressions reduces to R\'enyi case for
$r=\alpha$, and to the Tsallis case for $r=1$.
To the best of our knowledge, these relationships have not
previously been derived in the literature and can be used as a base
for future applications of supra-extensive entropy.

\section{Conclusion}\label{conc}

In this paper we considered generalized maximum entropy thermostatistics, under general type of internal energy constraints, for the class of strongly pseudo-additive (SPA) entropies which can be represented as an increasing continuous transformation $h$ of R\'enyi entropy. We developed a SPA-R\'enyi entropy duality principle by which the thermostatistics formalism of R\'enyi entropy $\Req$ is transformed in
the thermostatistics formalism of SPA entropy $h(\Req)$ and established the conditions for the function $h$ which preserve Legendre structure of thermodynamics when passing from one to another formalism. We considered the question what are the choices of function $h$ for which the Legendre structure is preserved and we shown that, in general, in the case of positive R\'enyi heat capacity, concavity of the function $h$ represents a sufficient condition for the equivalence. In special cases of the R\'enyi heat capacity equals to unity, the formalisms are shown to be equivalent for any choice of $h$, excluding the case that corresponds to Sharma-Mittal entropy \cite{Frank_Plastino_02} with the non-extenivity parameter $q=0$. In addition, we derived general entropy-constraint duality, by which the SPA (and R\'enyi) thermostatistics which corresponds to the linear constraints can be transformed into the one which corresponds to the escort constraint.

Thus, we established the equivalence between four different thermostatistics formalisms based on R\'enyi and SPA entropies coupled with linear and escort constraints. Using the equivalence, we provided the transformation formulas from one to another formalisms, we derived corresponding maximum entropy distributions, and we established new relationships between the corresponding thermodynamic potentials and temperatures. In this way we obtained a consistent framework that unifies and generalizes the results for wide class of entropies and constraints previously discussed in the literature, which is illustrated with several examples. As a special case, we derived the expressions for maximum entropy distributions and thermodynamic quantities for Sharma-Mittal entropy class \cite{Frank_Plastino_02}, which includes previously derived expressions for Tsallis, R\'enyi and Gaussian entropies. In addition, the results are applied for the derivation of maximum entropy distributions and thermostatistical relationships for supra-extensive entropy \cite{masi2005step}, which have not been considered so far. Presented framework could also be applied for thermostatistical analysis of another SPA entropies not considered here, such as Bekenstein-Hawking entropy \cite{czinner2015black} and super-exponential entropy \cite{jensen2018statistical}, as well as for more general group entropies \cite{tempesta2011group}, \cite{curado2016new} \cite{tempesta2016beyond}, \cite{Ilic_Stankovic_14a}, \cite{jizba2016q}, \cite{Ilic_Stankovic_17}. The framework can particularly be useful for the analysis of thermodynamic stability of SPA entropies \cite{Wada_02}, \cite{scarfone2006thermal} which will be discussed elsewhere.

\section*{Acknowledgements}

The first named author (V.M.I.) is supported by Ministry of Science and Technological Development, Republic of Serbia, Grant No. III044006 and Grant No. ON174026. The third named author (T.W.) is partially supported by Japan Society for the Promotion of Science (JSPS) Grants-in-Aid for Scientific Research (KAKENHI) Grant Number JP17K05341.

\providecommand{\noopsort}[1]{}\providecommand{\singleletter}[1]{#1}%

\end{document}